%% file: kheradmand-rosu-2018-tr.tex
\documentclass[sigconf,screen]{acmart}

\setcopyright{none}

\renewcommand\footnotetextcopyrightpermission[1]{} 
\usepackage{listings}
\usepackage{textcomp} 
\usepackage{microtype} 

\usepackage{tikz}
\usetikzlibrary{arrows}

\usepackage{booktabs}   
\usepackage{fancyvrb}

\input{macro.tex}
\input{header.tex}

\usepackage{xcolor}
\lstdefinelanguage{p4}{keywords={header_type,header,parser,action,table,control,register,field_list,counter,meter,metadata,fields,reads,actions,if,return,field_list_calculation,output_width,algorithm,input,calculated_field,verify,update}}

\begin{document}

\title{P4K: A Formal Semantics of P4 and Applications}

\author{Ali Kheradmand}
\affiliation{
  \institution{University of Illinois at Urbana-Champaign}
}
\email{kheradm2@illinois.edu}

\author{Grigore Rosu}
\affiliation{
  \institution{University of Illinois at Urbana-Champaign}
}
\email{grosu@illinois.edu}

\begin{abstract}
\input{sections/abstract}
\end{abstract}

\maketitle

\input{sections/introduction}
\input{sections/background}
\input{sections/p4k}
\input{sections/evaluation}
\input{sections/applications}
\input{sections/related_work}
\input{sections/conclusion}
\input{sections/acknowledgments}

\bibliography{bibliography}

\end{document}

%% file: macro.tex
\newcommand{\pft}{P4\textsubscript{14}\xspace}
\newcommand{\pst}{P4\textsubscript{16}\xspace}

\newcommand{\cellk}[1]{\texttt{#1}\xspace}
\newcommand{\pcode}[1]{\texttt{#1}\xspace}

%% file: header.tex
\PassOptionsToPackage{pdftex,usenames,dvipsnames,svgnames,x11names}{xcolor}
\PassOptionsToPackage{pdftex}{hyperref}
\usepackage[style=math]{k}

\newcommand{\reduceMulti}[2]{\hbox{%
  \begin{tikzpicture}[baseline=(top.south), 
                      inner xsep=0pt,
                      inner ysep=.3333ex,
                      minimum width=2em]
    \path
          node (top) [
            inner ysep=0.6ex
          ]{ $ \begin{array}{@{}c@{}}
                #1
               \end{array} $ \mathstrut}
          (top.south)
          node (bottom) [
            anchor=north,
          ] {
            $ \begin{array}{@{}c@{}}
              #2
            \end{array} $};
    \path[draw,thin,solid] let \p1 = (current bounding box.west),
                               \p2 = (current bounding box.east),
                               \p3 = (top.south)
                           in (\x1,\y3) -- (\x2,\y3);
    \path[fill] (top.south) ++(2pt,0) -- ++(-4pt,0) -- ++(2pt,-1.5pt) -- cycle;
  \end{tikzpicture}%
}}

\renewcommand{\reduce}[2]{\reduceMulti{#1}{#2}}

\renewcommand{\kall}[3][black]{\mall{#1}{#2}{#3}}

\newcommand{\syntaxContNewLine}[3][\defSort]{\par\indent\rulebox{%
  $\setlength{\syntaxlength}{\widthof{$\mathrel{::=}$}}%
  \setlength{\syntaxlength}{.5\syntaxlength}%
  \addtolength{\syntaxlength}{\widthof{\syntaxKeyword$#1$}}%
  \hspace{\syntaxlength}%
  \;\;\;\;\;\;\;{#2}$ \ifthenelse{\equal{#3}{}}{}{[#3]}%
  }
}

\renewcommand{\kall}[3][white]{\mall{black}{#2}{#3}}

\renewcommand{\kprefix}[3][white]{\mprefix{black}{#2}{#3}}

\usepackage{acronym}

\newcommand{\kequation}[2]{\begin{equation*}{\small#2}\end{equation*}}



%% file: sections/abstract.tex
Programmable packet processors and P4 as a programming language for such
devices have gained significant interest, because their flexibility
enables rapid development of a diverse set of applications that work at
line rate.
However, this flexibility, combined with the complexity of devices and
networks, increases the chance of introducing subtle bugs that are hard
to discover manually.
Worse, this is a domain where bugs can have catastrophic consequences,
yet formal analysis tools for P4 programs / networks are missing.

We argue that formal analysis tools must be based on a formal semantics
of the target language, rather than on its informal specification.
To this end, we provide an executable formal semantics of the P4 language
in the K framework.
Based on this semantics, K provides an interpreter and various analysis tools
including a symbolic model checker and a deductive program verifier for P4.

This paper overviews our formal K semantics of P4, as well as several
P4 language design issues that we found during our formalization process.
We also discuss some applications resulting from the tools provided by K for P4
programmers and network administrators as well as language designers and
compiler developers, such as detection of unportable code, state space
exploration of P4 programs and of networks, bug finding using symbolic
execution, data plane verification, program verification, and translation
validation.


%% file: sections/introduction.tex
\section{Introduction}
\label{sec:introduction}
As an increasingly important part of our life and society,
computer networks have grown significantly in size and complexity.
Traditionally, to handle the network scale, the networking hardware
has been hard coded with well-established network protocols
needed to run and manage the network.
However, doing so has the downside of not being able to cope with
the speed of innovation that is necessary to satisfy the diverse and
growing set of user demands, because the process of modifying networking
equipment tends to be slow and expensive.
This has ignited a line of research whose goal is to make networks
more programmable.


One of the most recent developments in this line of research, P4~\cite{p4}, is a
high level declarative programming language for programming packet processors.
P4 allows the developers to specify how a packet processor should process its incoming packets.
A P4 compiler then translates the P4 program into an instruction set understandable by the target hardware.
The examples of targets include software switches, high performance ASICs, FPGAs, and programmable NICs.

Since its introduction in 2014, P4 has attracted significant interest because
the flexibility that it provides enables rapid development of a diverse set of
applications that can potentially work at line rate, such as
In-Band Network Telemetry~\cite{int} and switch based implementation
of Paxos~\cite{p4paxos}.
However, this flexibility, combined with the complexity of networks and
networking hardware, increases the chance of introducing subtle
bugs that are very hard to discover manually, yet can have catastrophic effects,
from service disruptions to security vulnerabilities.

Even without P4, answering the simplest questions about the correctness of
a network (e.g., what kind of packets can reach node B from node A) has become
manually prohibitive when the scale and complexity of networks is taken into
account.
Subsequently, a large body of research has recently focused on automating the
process of network verification~\cite{hsa, netplumber, veriflow, deltanet}.
However, most of these works assume a simple fixed structure for the packet
processors and, as a result, may miss many details.
P4 makes manual verification even harder, if not impossible.
Consequently, there is a big need for automated tools to analyze P4 programs
or networks of nodes programmed using P4.

We adhere to \cite{kprover} that analysis tools for any programming language
must be based on the formal semantics of that language rather than on its
informal specification.
Informal semantics are subject to interpretation by different tool developers
and usually there is no guarantee that these interpretations are consistent
with the specification or with each other.
As shown in \cite{kprover}, state-of-the-art program analysis tools based on
informal language specifications ``prove'' incorrect properties or fail to
prove correct properties of programs due to their misinterpretation of the
semantics of the target programming language.
Moreover, the informal language specification itself might have problems, such
as ambiguities, inconsistencies, or even parts of the language not defined at
all.
This is particularly relevant for new languages, like P4, whose design has not
matured yet.

It is therefore important to develop a formal semantics for P4.
Furthermore, to build confidence in the adequacy of a formal semantics, we believe
it should be:
(1) \emph{executable}, so it can be rigorously tested against potentially
hundreds of programs;
(2) \emph{compact and human readable}, so it can be easily inspected and
ultimately trusted by everyone.
Finally it must be (3) \emph{modular}, so new language features can be
formalized without the need to change the previously formalized features.

To this end, we have developed P4K, an executable formal semantics
of P4 based on the official P4 language specification~\cite{p4v104}.
P4K faithfully formalizes all of the language features mentioned in the
P4 specification, with a few exceptions 
corresponding to features whose meaning was ambiguous or incorrect
or under specified and we did not find any satisfactory way to correct it.
We have reported some of these issues to the P4 language designers
\cite{p4-issue-neg, p4-issue-lpm, p4-issue-stateful-initial, p4-issue-direct-meters, p4-issue-variable-length-header, p4-issue-header-instance, p4-issue-push-pop, p4-issue-payload, p4-issue-update-verify-order}
and are working on a modified version of the specification~\cite{p4v105} addressing the issues.
We validated P4K by executing 40 test cases provided by one of the
official compiler front-ends of P4~\cite{p4c}, a manually crafted test
suite of 30 tests, and by formally analyzing several programs.

We chose the K framework ~\cite{k} for our P4 formalization effort.
It has several advantages that make it a suitable choice.
First, a language defined in K enjoys all three properties mentioned above.
Second, once a programming language semantics is given, K
automatically provides various tools for the language, including an
interpreter and a symbolic model checker, at no additional effort.
Finally, K has already been successfully used to formalize the semantics
of major programming languages such and C~\cite{kcc},
Java~\cite{kjava}, JavaScript~\cite{kjs}, etc.

The focus of this paper is the P4K formalization of P4, but we also
show how P4K and the tools provided by K can be used beyond just a reference
model for the language.
We discuss several applications useful for P4 programmers, language designers,
and compiler developers, such as:
detection of unportable code,
state space exploration of P4 programs and of networks,
bug finding using symbolic execution,
data plane verification,
deductive verification,
and translation validation.
Specifically, we make the following contributions:

\begin{itemize}
\item P4K: the most complete formal semantics of P4, based on the
official specification of \pft v. 1.0.4.
\item A collection of P4 formal analysis tools for the networking domain,
automatically derived from the semantics.
\end{itemize}

The paper is organized as follows.
Section~\ref{sec:background} overviews P4 and K,
as well as the challenges in defining a semantics for P4.
Section~\ref{sec:p4k} describes P4K, our K semantics of P4,
and discusses some of problems that we identified in the language specification.
Section~\ref{sec:evaluation} evaluates our semantics.
In Section~\ref{sec:applications} some of the applications of the semantics
are discussed.
Section~\ref{sec:related_work} reviews related work and
Section~\ref{sec:conclusion} concludes.


%% file: sections/background.tex
\section{Background and Challenges}
\label{sec:background}
Here we give background on P4 and K.
We also discuss some of the challenges that we faced in formalizing P4.

\subsection {Software Defined Networks}
\textit{Control plane} is the part of the network responsible for making packet
forwarding decisions by running computations (e.g. routing algorithms) based on
the network state.
\textit{Data plane} is the collection of forwarding devices
(or packet processors) that actually carry the network packets
and execute the forwarding decisions.
Traditionally, each device had its own vendor-provided control plane hard-coded
on the device.
The need for rapid innovation has sparked interest in Software Defined Networks (SDNs).
SDN is a modern architecture in which the control plane
is physically separated from the data plane.
In this architecture, one controller can \textit{program} a set
of forwarding devices through open, vendor-agnostic interfaces such as
OpenFlow~\cite{openflow}.

In OpenFlow, each device processes the packets
according to the contents of one or more \textit{flow tables}.
Each table will contain a set of \textit{flow entries}.
Abstractly, each entry is a \textit{(match, action)} tuple.
\textit{Match} provides values for specific fields in the packet header,
and \textit{action} denotes the action to be performed if the packet header
matches the respective values in match.
Possible actions include dropping, modifying, or forwarding the packet.
The controller programs the data plane through installation and modification
of flow entries.

OpenFlow assumes a fixed structure for the forwarding devices.
It explicitly specifies the set of protocol headers on which it operates,
the structure of the flow tables, the set of possible actions, etc.
Modification to any of these features requires an update to
the OpenFlow specification.
Over the course of 4 years since the initial version of OpenFlow, the number of
supported header fields in its specification has been more than
tripled~\cite{p4}.

\subsection{P4}
The limitations of OpenFlow and the need for expressiveness
has lead to the introduction of P4,
a high level declarative programming language for expressing the
behavior of packet processors.
In P4, one can program a custom parser for their own protocol header,
define flow tables with customized structure, define the control flow
between the tables, and define custom actions.
Hence, P4 allows the developers to specify how a packet processor
should process its incoming packets
(note, however, that P4 does \emph{not} provide a mechanism to
populate the flow table entries; this is done by the controller).
A P4 compiler then translates the P4 program into the
instruction set of the hardware of the packet processor
on which the program is installed (the \textit{target}).

We briefly describe the basic notions of the language here.
Section~\ref{sec:p4k} discusses the language construct in more details.
A P4 program specifies at least the following components~\cite{p4v104}.
\textit{Header types:} Each specifies the format (the set and size of fields) of
a custom header within a packet.
\textit{Instances:} Each is an instance of a header type.
\textit{Parser:} A state machine describing how the input packet
is parsed into header instances.
\textit{Tables:} Each specifies a set of fields that the table entries can
match on and a set of possible actions that can be taken.
\textit{Actions:} Each (compound action) is composed of a
 set of primitive actions which can modify packets and state.
\textit{Control flow:} Describes the custom conditional chaining of tables
within the packet processor's pipeline.

\begin{figure}[t]
\vspace*{-2ex}
  \centering
  \begin{lstlisting}[language=p4,basicstyle=\footnotesize\ttfamily]
  header_type h_t { fields { f1 : 8; f2 : 8; } }
  header h_t h1;
  parser start { extract(h1); return ingress; }
  action a(n) {
    modify_field(h1.f2, n);
    modify_field(standard_metadata.egress_spec, 1);}
  action b() {
    modify_field(standard_metadata.egress_spec, 2);}
  table t {
    reads { h1.f1 : exact;} actions { a; b; }}
  control ingress { apply(t); }
  \end{lstlisting}
\vspace*{-2ex}
  \caption{A very simple P4 program}
  \label{fig:simple-p4-program}
\vspace*{-4ex}
\end{figure}

For example, in Figure~\ref{fig:simple-p4-program}, \pcode{h\_t} is a header type
consisting of two 8 bit fields \pcode{f1} and \pcode{f2}. Header \pcode{h1} is
an instantiation of \pcode{h\_t}.
The parser consists of a single state \pcode{start} which \textit{extracts}
\pcode{h1} from the input packet.
There are two compound actions \pcode{a} and \pcode{b} in the program.
The actions use the \pcode{modify\_field} primitive action.
The entries in table \pcode{t} match on field \pcode{f1} in \pcode{h1} and if
applied, may call actions \pcode{a} or \pcode{b}.
An entry calling \pcode{a} must provide a value for \pcode{n}.
The \pcode{ingress} pipeline in this program only consists of applying table
\pcode{t}.

\begin{figure}[t]
\vspace*{-2ex}
  \centering
  \includegraphics[width=\columnwidth]{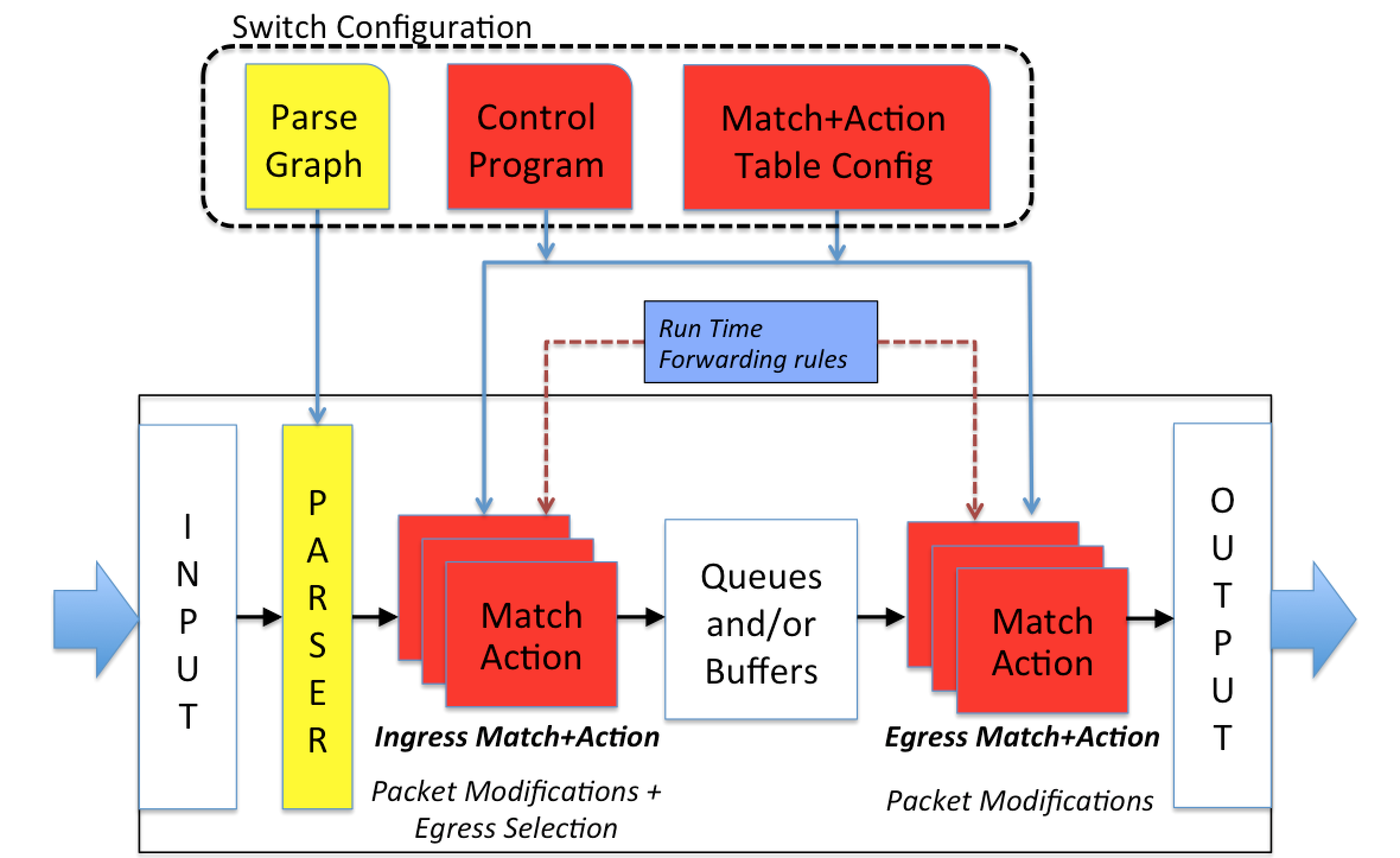}
\vspace*{-4ex}
  \caption{\pft Abstract Forwarding Model~\cite{p4v104}}
  \label{fig:afm}
\vspace*{-3ex}
\end{figure}

P4 programs operate according to the abstract forwarding model
illustrated in Figure~\ref{fig:afm}.
For each packet, the parser produces a \textit{parsed representation} comprised of
header instances and sends it to the \textit{ingress} match+action pipeline.
Ingress, among other things, may set the \textit{egress specification} which
determines the output port(s).
Ingress then submits the packet to the queuing mechanism the specification
of which is out of the scope of P4.
The packet may also go through the optional \textit{egress} pipeline which
may make further modifications to the packet.
Finally (if not dropped) the parsed representation will be \textit{deparsed}
(i.e serialized) into the packet which will be sent out.
P4 also supports re-circulation (looping packets inside the device)
and cloning of packets.

The P4 Language Consortium (\url{http://p4.org}) provides the official
specification of the language, as well as various other tools including
compiler front ends, software interpreters, runtime and
testing infrastructure, etc.
There are two versions of P4 in current use, \pft and \pst.
\pst has been released by the consortium in May 2017~\cite{p416}
and it is much simpler and cleaner than \pft, at the cost of
deliberately breaking backwards compatibility with \pft.
There are, however, important \pft program which have not
been translated to \pst, and, indeed, the \pft-to-\pst translator
provided by the consortium is not
semantics-preserving~\cite{p4-14-16-bmv2-header-stacks}.
Ideally, we would like to prove the translator correct, but for
that we need formal semantics of both \pft and \pst.
In this paper we only discuss our formal semantics of \pft, leaving
that of \pst and the correctness of the translator as future
work.
Throughout the paper, we refer to \pft simply as P4.




\subsection{Challenges in Formalizing P4}
\label{sec:p4-challenges}
P4 has several characteristics that make it a challenging target for formalization.
Here we discuss some of this challenges and the way we dealt with them.

\textit{Unstable language:}
P4 is a relatively young language and it takes time for the community to
reach consensus on its design.
When we started, the only publicly available version of P4 was \pft v. 1.1.0.
That version soon was deprecated and replaced with v. 1.0.3 which we
initially used to develop our semantics.
In the middle of our formalization effort, \pst v. 1.0.0 as well as a minor
revision of \pft (v. 1.0.4) were released.
Thanks to K's support for modular definitions and reuse, we were able to
rapidly adapt to changes and finalize our semantics according to \pft v. 1.0.4.
Through continuous discussions with the P4 designers, we hope to help them
reach more stable versions sooner.


\textit{Imprecise specification:}
Since P4 is a newly developed language its specification is not free of
problems.
There are many inconsistencies and corner cases which are not discussed
(Section~\ref{sec:p4k}).
One of the important contributions of this paper is the identification
of these problems through rigorous formalization.
We have reported some of the problems to the community.
We are also working on a modified version of the specification~\cite{p4v105}
addressing the issues we found.

\textit{No comprehensive test suite:}
Similar formalization efforts for other languages
(e.g.~\cite{kcc,kjs,kevm}) rely heavily on official test suites.
Unfortunately, there is no official test suite for P4.
To alleviate this problem, in addition to testing our semantics against
a test suite we obtained from a P4 compiler, which only covers about half of
our semantic rules (Section~\ref{sec:evaluation}), we hand-crafted a test suite
that gives a complete coverage of our semantic rules.

\textit{Unconventional input:}
The input to a P4 program is different from that of conventional programming
languages.
P4 has two sources of input.
One is the stream of incoming packets that the device running the P4 program
needs to process.
The other is the table entries and configurations that are installed
by the controller at runtime.
The mechanism by which the controller interacts with the target at runtime
is device-specific and is therefore out of the scope of the language
specification.
Still, to be able to execute and analyze P4 programs, for the target-specific
language features we tried to provide the most unrestricted executable semantics;
for example, if the order of some operations was unspecified then we chose a
non-deterministic semantics, so we can still explore the entire state-space of
behaviors using the K tools.

We also grouped most of the target specific semantic rules
in a separate semantic module.
This way, the semantics is parametric on the target specific details.
One can provide a new target specific module to change the
target specific behavior, without the need to touch the rest of the semantics.
We have already used this feature when we were testing our semantics against
the p4c test suit as it contained target specific features and assumptions (Section~\ref{sec:evaluation}).
%
%

\vspace{-1ex}

\subsection{The K Framework}
K~\cite{k} is a programming language semantics engineering framework based
on term rewriting.
Its underlying philosophy is that tools for a language can and
should be automatically derived from the formal semantics of that language.
Indeed, K provides an actively growing set of language-independent tools,
i.e., tools which are not specific to any language but apply to any
language which has a K formal semantics.
These include a parser, an interpreter, a symbolic model checker,
a sound and (relatively) complete deductive program verifier, and,
more recently, a cross language program equivalence checker and a
semantic-based compiler.
Some of the tools are useful during the formalization process itself,
the most important of which is the interpreter.
Using the interpreter, the semantics can be tested against potentially many
programs to gain confidence in its correctness.

To define a programming language in K, one needs to define its syntax
and its semantics.
Syntax is defined using BNF grammars annotated with semantic attributes.
Semantics is given using rewrite rules (also called semantic rules)
over configurations.
A configuration is a set of potentially nested cells that hold the
program and its context.
Each cell contains a piece of semantic information of the input program
such as the its state, environment, storage, etc.
Semantic rules are transitions between configurations:
if parts of the configuration match its left hand side,
rewrite those parts as specified by the right hand side.
For example, this is a rule taken
from P4K (modified for presentation) concerning reading the value of
field \pcode{F} from instance \pcode{I}:
\vspace*{-2ex}

{\footnotesize
\kequation{ref}{
\kprefix{k}{\reduce{\variable{I}.\variable{F}}{\variable{V}}} \mathrel{}
    \kall{instance}{
      \kall{name}{\variable{I}} \mathrel{}
      \kall{valid}{\variable{true}} \mathrel{}
      \kall{fieldVals}{\ellipses\variable{F}\mapsto\variable{V}\ellipses} \mathrel{}
      \ellipses
    }
}
}
The contents inside each matching pair of angle braces constitutes a cell,
with the cell name as subscript. 
The \cellk{k} cell contains the list of computations to be executed.
The fragment of computation at the front of the list (the left most) is executed first.
There are multiple \cellk{instance} cells each corresponding to a
header instance.
\cellk{name} contains the name of the instance and \cellk{valid}
keeps its validity state.
\cellk{fieldVals} is a map from each field name to the value stored
in the field in the given instance.
The ellipsis are part of the syntax of K and denote contents irrelevant to the rule.
The horizontal line denotes a rewrite.
If the configuration matches the pattern, the part of the configuration above
the line will be replaced by the content below the line.
The rest of the configuration remain intact.
A rule may contain multiple rewrites at different positions of
the configuration.
In that case, all rewrites will be applied in one step.

This example illustrates two properties of K that makes it suitable
for giving semantics specially to evolving programming languages like P4.
First, note that the actual configuration contains many more cells
and each cell may contain multiple elements, but the rule only
mentions the cells that are relevant.
The \textit{configuration abstraction} feature of K automatically infers what the rest of cells should be.
Second, note that rewrites are local. 
There is no need to rewrite the whole configuration.
These two features make K rules succinct and human readable.
More importantly, they enable modular development of the semantics:
if the language specification adds or modifies a language feature
the rules irrelevant to that feature do not need to be modified.




%% file: sections/p4k.tex
\section{P4K}
\label{sec:p4k}
P4K is the most complete executable formal semantics of \pft.
It is based on the latest official language specification
(v. 1.0.4~\cite{p4v104}) and on discussions with the language designers.
Our work is open source and is available online~\cite{p4k}.
%
The formalization process took 6 months to complete by a PhD student
with some familiarity with the K framework.
Most of the time was spent learning K and understanding the details
of the P4 specification, including its problems.
P4K contains more than 100 cells in the configuration, 400 semantic rules,
200 syntax productions, and 2000 lines.

\subsection{Syntax}
The language specification provides a BNF grammar, whose conversion to K
was straightforward.
We mostly copy-pasted the grammar and made a few minor modifications
to make it compatible with K.

During this process, in addition to minor problems, we
identified~\cite{p4-issue-neg} an ambiguity in the syntax between the minus sign
in a constant value (for specifying negative constant values) and the unary
negation operator.
This ambiguity has important semantic effects.
In P4, all the field values have a bit width associated with them.
According to the specification
``For positive [constant] values the inferred width is the smallest number of bits required to contain the value''.
Also ``For negative [constant] values the inferred width is one more than the
smallest number of bits required to contain the positive value''~\cite{p4v104}.
So for example $-5$ interpreted as a negative constant would yield a 4 bit
value while if interpreted as negation of a positive constant would yield
a 3 bit result.
Used in an expression with other operators, this difference may affect whether
the expression overflows or not, which subsequently may affect the final result.

\subsection{Configuration}

\input{configuration.tex}

The configuration contains more than 100 cells.
Figure~\ref{fig:config} shows part of it, featuring
more important cells.
All of the language constructs including headers, instances, parser states,
actions, tables, control flows, etc have respective cells in the configuration
containing their static information and/or runtime state. 
For example the \cellk{tables} cell will contain a set of \cellk{table} cells
(``*'' denotes multiple cells with the same name).
Each of the \cellk{table} cells contains a table's static information such as
its \cellk{name}, the fields to match (\cellk{reads}), and possible
actions (\cellk{acts}).
It also contains runtime information such as the
\cellk{entries} installed in the table.

Some cells contain the execution context.
For example during the execution of an action, the \cellk{stackFrame}
cell holds a stack of maps from each formal parameter of the executing
action to the respective argument values passed to the action.

The cells \cellk{in} and \cellk{out} contain the input and output packet stream from/to all ports respectively.
\cellk{packetin} contains the current packet being processed and \cellk{packout} contains the packet being serialized.

Cells are populated or modified by processing the input P4 program before
execution, during the initialization, or during the execution
as discussed next.

\subsection{Semantics}
After parsing,
the P4 program populates the \cellk{k} cell and is
executed with the semantics rules.

\subsubsection{Execution Phases}
The rules describe the P4 program execution, in three phases.

\textbf{Preprocessing:}
In this phase, P4K iterates over all the declarations in the input P4
program (in the \cellk{k} cell), creating and populating the corresponding
cells and preparing the configuration for execution.
In some cases auxiliary
information is pre-computed for the execution phase.
An important such computation is the inference
of the order of packet headers for deparsing.
Details will be discussed in Section~\ref{sec:p4k-semantics}.


\textbf{Initialization:}
There is an optional initialization phase after
preprocessing.
It is used primarily to prepopulate the tables and packet buffers
before the execution in certain analysis such as symbolic execution.
The tables and packet buffers can also be populated at runtime in normal execution.

\textbf{Runtime:}
The actual execution of a P4 program happens during this phase.
It implements the abstract forwarding model.
Packets are taken from the input packet stream and processed using the entries
installed in the match+action tables by going through the ingress
and potentially the egress pipelines.
The output packets are appended to the output packet stream.
This phase never terminates.

\vspace*{-2ex}

\subsubsection{Language Constructs and Semantics}
\label{sec:p4k-semantics}
We briefly describe the language constructs and primarily
focus on interesting findings and relevant semantics.

\textbf{Header types}:
Each header type is a named declaration that includes an ordered list
of fields and their attributes (e.g. field width and signedness).
P4 also allows declaration of variable length headers.
During our formalization, we found corner cases
(e.g.~\cite{p4-issue-variable-length-header}) in which the semantics of such
headers are not completely clear.

\textbf{Instances}:
Instances may be referenced in various runtime stages including
parsing, table matching, and action execution.
Some instances, called \textit{header instances}
(although the naming is not consistent throughout
the specification~\cite{p4-issue-header-instance}),
keep the parsed representation of the respective packet headers
(i.e., the packet header is extracted into the header instance).
Other instances, called \textit{metadata}, keep arbitrary per packet
state throughout the pipeline.
For example \pcode{h1} in Figure~\ref{fig:simple-p4-program} is a header instance and
\pcode{meta} in Figure~\ref{fig:load-balancer} is a metadata.
In our semantics, both types of instances are kept as \cellk{instance} cells,
distinguished by their \cellk{metadata} cells (Figure~\ref{fig:config}).

It is also possible to declare fixed size, one dimensional array instances
(called \textit{header stacks}), as sequences of adjacent headers (e.g. to
support MPLS~\cite{mpls}).
We keep array elements as separate header instances,
with special names that include their index.
Otherwise, the elements are treated same as other instances.

Header instances are invalid (uninitialized) until validated in parsing
or by specific primitive actions in match+action processing.
According to the specification, reading an invalid header results in
an undefined value, whose behavior is target dependent.
We model this using a special value \pcode{@undef}.
Use of \pcode{@undef} in an expression or action call
causes the execution to get stuck by default.
We use this feature to detect unportable code
(Section~\ref{sec:undefined-behavior}).

\textbf{Hash generators}:
The ability to calculate a hash value for a stream of bytes
has various uses in networking.
P4 provides the ability to declare hash generators
(called \textit{field list calculation}s).
The developer provides a list of values (declared using a
\textit{field list} struct) and selects a hash generation algorithm.
The hash generator computes the hash of the bitstream generated from the list.
In the example below \pcode{ipv4\_checksum} is a hash generator for the
\pcode{ipv4\_checksum\_list} field list (declaration omitted) with the IPv4 checksum algorithm
(\pcode{csum16}).
\vspace{-1ex}
\begin{lstlisting}[language=p4,basicstyle=\footnotesize\ttfamily]
field_list_calculation ipv4_checksum {
    input {ipv4_checksum_list;}
    algorithm : csum16; output_width : 16; }
\end{lstlisting}
\vspace{-1ex}
The language specification
identifies a set of well known hash generation
algorithms (e.g., IPv4 checksum and CRC).
In our semantics, we treat hash generation as a black box; K allows us to
``hook'' library function calls that implement the desired functionality.
It is possible to also directly specify the algorithms using K rules inside
the semantics, but we did not find any compelling reason to do so.

We found a problem~\cite{p4-issue-payload} with the specification during the
formalization of field lists.
Each element of a field list can refer to
a field in an instance,
an instance itself (when all the fields in that instance are used),
another field list (when all the fields identified by that list are used),
a constant value,
or the keyword \pcode{payload}.
According to the specification
``payload indicates that the contents of the packet
following the header of the previously mentioned field is included
in the field list''~\cite{p4v104}.
However, ``previously mentioned field'' is ambiguous.
For instance, below it is not clear if payload refers to f1 or f2.
\vspace{-1ex}
\begin{lstlisting}[language=p4,basicstyle=\footnotesize\ttfamily]
field_list fl1 { h.f1; }
field_list fl2 { h.f2; fl1; payload; }
\end{lstlisting}
\vspace{-1ex}
Thus we do not provide semantics for payload.
\pst has replaced field lists with a C-like \textit{struct} construct,
disallowing the payload keyword.


\textbf{Checksums}:
A field in a header instance can be declared to be a \textit{calculated field},
indicating that it carries a checksum.
The developer provides a hash generator for verification of the
checksum at the end of parsing,
and/or an update of the checksum during deparsing.
For example, below, the field \pcode{hdrChecksum} in the header instance
\pcode{ipv4} is declared to be a calculated field which uses the hash generator
\pcode{ipv4\_checksum} for its verification and update.
\vspace{-1ex}
\begin{lstlisting}[language=p4,basicstyle=\footnotesize\ttfamily]
calculated_field ipv4.hdrChecksum  {
    verify ipv4_checksum; update ipv4_checksum;}
\end{lstlisting}
\vspace{-1ex}
The P4 specification leaves undefined the order in which the
calculated fields must be updated or verified.
For verification, the order can matter depending on the target.
For update, the order can matter in cases where the
field list calculation of a calculated field includes another
calculated field.
After discussing~\cite{p4-issue-update-verify-order} with the
language designers, to obtain the most general behavior, we decided to
choose a non-deterministic order for update and verify.
K provides a search tool which one can use to explore all
possible non-deterministic outcomes to check whether they differ.
(Section~\ref{sec:model-checking}).


\textbf{Parser}:
The user can define a parser to deserialize the input packet
into header instances (the \textit{parsed representation}).
The parser is defined as a state machine.
In each state, it is possible to
\textit{extract} header instances (i.e., copy the data from the packet at the
current offset into respective field values for the given instances) and
to modify metadata.
Then, it is possible to conditionally transition to another state,
to end the parsing, or to throw an (explicit) exception.
For example, in state \pcode{parse\_ethernet} in Figure~\ref{fig:basic-routing},
after extracting the \pcode{ethernet} header, based on the value of the
\pcode{etherType} field, the parser may transition to the parser state
\pcode{parse\_ipv4} or end the parsing and start the ingress pipeline.


\textbf{Exception Handlers}:
P4 allows us to declare exception handlers for implicit
or explicit parser exceptions.
In case an exception occurs, parsing is terminated and the
relevant handler is invoked.
Each handler can either modify metadata and continue to ingress
or immediately drop the packet.
There is a default handler that drops the packet.
For example in Figure~\ref{fig:simple-p4-program} if a packet is too short for
the extraction of \pcode{h1}, an implicit exception is thrown and the default
handler drops the packet.

\textbf{Deparsing}:
This is the opposite of parsing.
At egress, the (potentially modified) valid header instances
are serialized into a stream of bytes to be sent.
An important question in deparsing is the order in which
the header instances should be serialized.
The parsing order is not enough to find the deparse order, since
header instances might also be added (validated) or removed (invalidated)
in the match+action pipeline.
According to the specification
``[A]ny format which should be generated on egress should be
represented by the parser used on ingress''~\cite{p4v104} and the
order of deparsing should be inferred from the parse graph.

If the parse graph is acyclic, a topological order can be used as
the deparsing order.
However, in general the graph might be cyclic as there may be
recursion in parsing.
While in simple cases an order can still be inferred
(e.g., cases where recursion is only used for the extraction of header stacks),
there are cases in which a meaningful order can not be inferred.
This is a well known problem~\cite{p4v110}.
In our semantics, we support simple cases of cyclic parse graphs.
All of the practical examples we have seen so far can be handled
by our semantics.

\pst has switched to an approach in which
the deparse order is explicitly defined by the programmer.

\textbf{Stateful Elements}:
P4 supports \textit{stateful} language constructs that can
hold state for longer than one packet, as opposed to per packet
state in instances.
\textit{Counters} count packets or bytes,
\textit{meters} measure data rates, and
\textit{registers} are general purpose stateful elements.
The declaration of each of these elements creates an array of memory
units.
The units may be \textit{direct}ly bound to the table entries.
In that case, the (counter and meter) units will automatically be
updated when the corresponding entry is matched in match+action.
The units may alternatively be \textit{static}; then they should
explicitly be accessed or updated via special primitive actions.
For example, in Figure~\ref{fig:load-balancer}, \pcode{reg} is a static register
with a single 8 bit memory cell.

In our definition we unified all these elements as instances of the
\cellk{stateful} cell (Figure~\ref{fig:config}) which can be accessed like registers.
Other operations are defined as functions which read and manipulate
the registers.
Each \cellk{stateful} cell in the configuration has a map from an
index to a value.
The index is either a table entry id (for direct)
or an array index (for static).
The mechanism of updating meters is target specific
(not part of the language specification).
Subsequently we do not perform any action in case a meter is updated.
If needed, one can add a mechanism in our target specific module.

The specification does not specify~\cite{p4-issue-stateful-initial}
the initial value of the stateful elements.
It is sensible to assume that the initial value of counters is 0,
and similarly for meters.
For registers, by default we initialize the registers to \pcode{@undef}.
%
Moreover, the specification is inconsistent~\cite{p4-issue-direct-meters}
about whether direct meters are allowed to be explicitly
updated by table actions.
To be consistent with counters and registers, we assume they are allowed.

Finally, if multiple counters/meters are directly bound to the
same table, the specification does not state~\cite{p4-issue-update-verify-order}
the order in which the elements must be updated when an entry in that
table is matched.
The order can affect the outcome in multi-threaded packet
processors (Section~\ref{sec:concurrency})
as there may be data races over stateful elements.
Again, we choose a non-deterministic order for updating
the counters/meters, so we can systematically
explore it using K's search.  


\textbf{Actions}:
\textit{Compound} actions are user defined imperative functions that
can take arguments and if called, perform a sequence of calls to other
compound actions or built-in \textit{primitive actions}.
Primitive actions provide various functionality including
arithmetic, addition/removal/modification of instances and header stacks,
access/modification of stateful elements, cloning, re-circulation, dropping the packet, etc.
Actions are executed as a result of table matches.

We formalized all the primitive actions
(see Section~\ref{sec:limitation} for limitations on clone primitive actions).
The specification does not specify the behavior of some corner cases, such as
shift with negative shift amount.
We intentionally do not provide semantics for such cases to detect
unportable code.


The previous version of the specification~\cite{p4v103}
stated that all primitive actions resulting from a table match
execute in parallel, making it unclear what the meaning of
the following:
\vspace{-2ex}
\begin{lstlisting}[language=p4,basicstyle=\footnotesize\ttfamily]
action a() {
  modify_field(h.f, 1); modify_field(h.f, 2); }
\end{lstlisting}
\vspace{-1ex}
\pcode{modify\_field(f,v)} is a primitive action that updates
field \pcode{f} with \pcode{v}.
The latest revision (1.0.4) switched to sequential
semantics, so we do not have to deal with this case anymore.

\textbf{Tables}:
Tables will be populated at runtime by the controller.
Each entry provides values for the fields that are specified in the declaration,
an action that should be executed if the entry is selected, and arguments to be
passed to the action.

The interaction mechanism between the controller and P4 target is
out of the scope of the specification.
Hence, the answer to questions such as what happens
if a table is modified by the controller while it is being applied
on a packet is target dependent.
We currently assume that modification and application of the same table
are mutually exclusive.

P4 provides various matching modes per each field.
For example, \textit{exact} matches exact numbers
and \textit{ternary} matches ternary bit vectors.
It is also possible to associate priorities with table entries.
In case more than one table entry is applicable, the rule with the
highest priority will be selected.
\textit{Longest Prefix Match (LPM)} is a special kind of ternary
match useful for IP prefixes.
The specification specifies how the relative priority of an entry
with LPM match can be inferred bases on the corresponding match
value of the entry.
However, it does not specify how the priority should be decided
in cases where there are more than one field with
LPM match type~\cite{p4-issue-lpm}.
We assume all entries have explicit (unique) priorities regardless
of their match types.
We keep the entries sorted in their descending order of priority.
To apply a table on a packet, we iterate over the entries in order
and select the first matching entry.

\textbf{Control Flow}:
User defines the order and the conditions under which various
tables are applied to a packet using \textit{control function}s.
The body of a control function is a control block
consisting of a sequence of control statements.
A statement might
apply a table,
call a control function, or
conditionally select a control block.
Ingress is a special control function that is automatically called after
(successful) parsing.
Egress is another (optional) special control function.
If defined, it will automatically be called when the queuing mechanism takes the packet
to be sent out.


\textbf{Other contstructs}:
We omit the discussion of \textit{value sets}, \textit{action profiles}, and
\textit{action selectors} as well as many details of the discussed constructs.
Interested readers can refer to the semantics~\cite{p4k} for more details.

\vspace*{-1ex}

\subsection{Concurrency Model}
\label{sec:concurrency}
Real world high performance packet processors have multiple threads of execution.
The specification is silent about the concurrency model.
As a result, what constitutes a thread depends on the target hardware.
In our semantics, we support a multithreading model in which
each thread individually does all of what a single threaded program does by
addition of a few more cells and rules\footnote{In all of our experiments we used only a single thread for each P4
program.
Throughout the paper we assume executions are single threaded.}
(similar to Section~\ref{sec:network-semantics}).
The input/out packet streams, the tables, and the stateful elements constitute
the shared memory between the threads.


\subsection{Limitations}
\label{sec:limitation}
P4 provides four primitive actions for cloning a packet under process from
ingress/egress to ingress/egress.
The actions that clone a packet into the egress put the clones
in the queue between ingress and egress pipelines.
Since we currently do not model the ingress and egress pipelines as
separate threads, we only support a single packet in the queue between the two.
Therefore, we do not directly support clone into egress.
Instead, we treat such clones as new incoming packets with auxiliary flags to
skip the ingress pipeline.



\subsection{Network Semantics}
\label{sec:network-semantics}
It is useful to be able to simulate or analyze a network
of P4 programs rather than just a single program
(Sections~\ref{sec:dataplane-verification} and~\ref{sec:model-checking}).
In order to do so, we need the semantics of the network.
Thanks to the modularity of K, we easily modeled the semantics of a
P4 network without changing the P4 language semantics.
We only needed to add a few more cells and preprocessing rules.
We added a root \cellk{nodes} cell containing multiple
\cellk{node} cells each containing the configuration of
a P4 program plus a \cellk{nodeId} cell.
We also added a \cellk{topology} cell which holds the
connection between the nodes.

To model the network links, we added a single rule that takes a packet from
the end of the output stream of one node and puts it at the beginning of
the input stream of the node it is connected to.
If needed, one can also model packet loss in the links
by a single additional rule.
Note that here we have multiple threads of concurrent execution,
whose interleaving is non-deterministic.
The thread interleaving space can be explored using the K search
mode (Section~\ref{sec:model-checking}).







%% file: configuration.tex
\begin{figure*}[t]
  \centering
  \renewcommand{\dotCt}[1]{\scriptstyle\textit{#1}}
  \newcommand{\id}{\scriptstyle\textit{Id}}
  \newcommand{\bool}{\scriptstyle\textit{Bool}}
  \newcommand{\kss}{\scriptstyle\textit{K}}
  \newcommand{\listss}{\scriptstyle\textit{List}}
  \newcommand{\xlistss}[1]{\scriptstyle\textit{List}_{\textit{#1}}}
  \newcommand{\xsetss}[1]{\scriptstyle\textit{Set}_{\textit{#1}}}
  \newcommand{\oid}{\scriptstyle\textit{ID}_{\sf obj}}
  \newcommand{\eid}{\scriptstyle\textit{ID}_{\sf env}}
  \newcommand{\obj}{\scriptstyle\textit{Var} \;\mapsto\; \textit{Val}_\textit{Prop}}
  \newcommand{\env}{\scriptstyle\textit{Var} \;\mapsto\; \textit{Val}_\textit{Env}}
  \newcommand{\activeStack}{\scriptstyle\textit{List}_{\sf running}}
  \newcommand{\excStack}{\scriptstyle\textit{List}_{\sf ctrl}}
  \newcommand{\pseudoStack}{\scriptstyle\textit{List}}
$
\kall{T}{
  \begin{array}{@{}c@{}}
  \kall{k}{\kss} \mathrel{}
  \kall{headers}{\dotCt{...}} \mathrel{}
  \kall{actions}{\dotCt{...}} \mathrel{}
  \kall{controlFlows}{\dotCt{...}} \mathrel{}
  \kall{parserStates}{\dotCt{...}} \mathrel{}
  \ellipses \mathrel{}
  \kall{tables}{
    \kall{table*}{
      \kall{name}{\id} \mathrel{}
      \kall{reads}{\xsetss{Fld}} \mathrel{}
      \kall{acts}{\xsetss{Act}} \mathrel{}
      \kall{entries}{\xlistss{Ent}} \mathrel{}
    }
  }
  \\ \mathrel{}
  \kall{instances}{
    \kall{instance*}{
      \kall{name}{\id} \mathrel{}
      \kall{metadata}{\bool} \mathrel{}
      \kall{valid}{\bool} \mathrel{}
      \kall{fieldVals}{\id \;\mapsto\; Val} \mathrel{}
    }
  }
  \kall{statefuls}{
    \kall{stateful*}{
      \kall{name}{\id} \mathrel{}
      \kall{vals}{\id \;\mapsto\; Val} \mathrel{}
    }
  }
  \\ \mathrel{}
  \kall{ctx}{
    \kall{stackFrame}{\xlistss{Map}} \mathrel{}
  }
  \kall{parser}{
    \kall{dporder}{\xlistss{Id}} \mathrel{}
    \kall{pctx}{
      \kall{index}{\dotCt{Int}} \mathrel{}
    }
  }
  \kall{packetin}{\dotCt{Pkt}} \mathrel{}
  \kall{packetout}{\dotCt{Pkt}} \mathrel{}
  \kall{buffer}{
    \kall{in}{\xlistss{Pkt}}
    \kall{out}{\xlistss{Pkt}}
  }
  \end{array}
}
$
  \caption{Part of the P4K configuration. The ellipsis symbols indicate omitted cells.}
  \label{fig:config}
\vspace*{-2ex}
\end{figure*}

%% file: sections/evaluation.tex
\section{Evaluation}
\label{sec:evaluation}

K provides us with an interpreter derived automatically
from the semantics, enabling us to test our semantics.
Official conformance test suits are an ideal target for
testing executable semantics.
Unfortunately, P4 does not have such a test suite.
A new official P4 compiler front end (p4c~\cite{p4c}) has a limited
set of tests for \pft, which we used in our evaluation.

Generally, it is non-trivial to port tests across different implementations
of P4, as its IO is not specified (Section~\ref{sec:p4-challenges}).
Fortunately, the p4c tests were easy to adopt.
Each test, along with the P4 program under test, contains an STF file.
The file describes table entries, input packets, and the expected output packets.
We systematically converted the STF files into our test format.

The suite contains tests with minor issues including use of deprecated
syntax, unspecified constructs, or unspecified primitive actions.
We fixed the issues by slightly modifying to the corresponding P4 programs or
test files, and implementing the primitive actions in a target specific
module for the tests.
Moreover, the tests assume undefined\footnote{According to the specification
egress specification is undefined unless set explicitly.
We model this using the \pcode{@undef} value.}
egress specification leads to packet drop.
The specification does not specify the behavior in this case,
so it is target dependent.
In our semantics, by default the execution gets stuck in such cases.
In our target dependent module for the tests, we added a rule to
drop the packet in such cases.

The tests also helped us identify a few problems in P4K.
For example, we found that we had misunderstood the semantics of a primitive
action (\texttt{pop}).
Note that \texttt{push} and \texttt{pop} have rather an unusual
semantics in P4~\cite{p4-issue-push-pop}.

After fixing the problems with the tests and our semantics, P4K
passed 39 out of the 40 test.
The failing test\footnote{Namely \texttt{parser\_dc\_full.stf}.}
has multiple inferable deparsing orders.
The order chosen in our execution
happens to be different from the order the test expects.
We verified that both orders are possible.

Inspired by~\cite{kjs}, we measured the percentage of the semantics rules
exercised by the tests (the \textit{semantic coverage} of the tests).
The tests cover under 54\% of the semantics and miss many of
the semantic features.
We have also manually developed 30 tests during our formalization process.
Together, these 70 tests cover almost all the semantic rules.

Each test took 19.5s ($\pm$ 3.2s) on average with the maximum of
125s.\footnote{All experiments are run on a machine with
Intel Xenon CPU ES-1660 3.30GHz and 32GB DDR3 1333MHz RAM.}
We note that approx. 10s out of this time is the
startup time of K and is not related to execution.
We also note that K has multiple backends.
We use an open source backend~\cite{kprover} which is
relatively very poor in terms of performance.
We expect the runtime to improve by orders of magnitude on
performant commercial backends (e.g.~\cite{rv-match}).




%% file: sections/applications.tex
\section{Applications}
\label{sec:applications}
Besides defining a formal semantics for P4 and thus helping make the P4
specification more precise, a secondary objective of our effort was to
make use of the various tools that K provides.
We demonstrate how the tools can be useful for the P4 developers and network
administrators, as well as for the P4 language designers and compiler
developers.

\subsection{Detecting Unportable Code}
\label{sec:undefined-behavior}

As seen above, in some cases the P4 specification does not provide the
expected behavior of the program.
P4 programs exhibiting such unspecified behavior may not be portable
among different targets and compilers.
It is not wise to solely rely on the expertise of P4 developers in the low
level details of the specification to check if their code is portable.
It is desirable to have tools that automate this check.
For simple cases, such behavior may be detectable by syntactic checks.
In general, unspecified behavior may depend on the input.

By default, we do not provide semantics for cases which are not covered by
the language specification.
If the execution of a program reaches a point with unspecified behavior,
the execution gets stuck.
Avoiding over-specification therefore allows us to check for unspecified
behavior in P4 programs.
This is done simply by running the program and checking whether it reaches
a state in which it gets stuck or not.
The check can be performed using either concrete or symbolic inputs.
We show a symbolic example in Section~\ref{sec:sym-exec}.

To tune the semantics for a specific target, one can provide custom
semantics for cases with unspecified behavior in the target specific module.

\subsection{State Space Exploration}
\label{sec:model-checking}
K provides a \textit{search} execution mode which allows us to explore
all possible execution traces when non-determinism is present.
In K, non-determinism occurs when more that one rewrite rule is applicable,
or the same rule is applicable at multiple positions in the configuration.
In normal execution mode, only one of the applicable rules is
(non-deterministically) selected.
In the search mode, all the applicable rules are explored.
Moreover, the user can explicitly control the points
in which non-determinism is explored.
This allows one to focus on exploration of one or more specific sources
of non-determinism and ignore the rest.

There are two sources of non-determinism in P4K.
The first is due to our approach to model the most
general behavior.
Examples are order of deparsing, order of update and/or verification of
calculated fields, and order of update of direct stateful elements.
The second is due to the existence of multiple threads of execution.
These include the threads of execution inside a single P4 program, as
well as the execution of multiple nodes in a P4 network.
Both sources can be explored using the search mode.
We have already shown in Section~\ref{sec:evaluation}
how exploration of the order of deparsing can be useful.
The benefits of exhaustive analysis of thread interleavings in
concurrency analysis are well known.

\subsection{Symbolic Execution}
\label{sec:sym-exec}
K allows the configuration to be symbolic --
i.e., to contain mathematical variables and logical constraints over them.
During execution with a symbolic configuration,
K accumulates and checks (using Z3~\cite{z3}) all
the logical constraints over the execution path
-- i.e the conditions under which the rules are applicable to respective states.
Under the hood, there is no difference between symbolic and concrete execution.
Symbolic execution powers some of the other K tools such as the
program verifier (Section~\ref{sec:deductive-verification}) and
the equivalence checker (Section~\ref{sec:translation-validation}).
It can also be useful on its own, say, to search for bugs in P4 programs
and data planes.

\subsubsection{Search for Bugs}
To illustrate one application, we choose a community provided sample P4 program
which defines a very basic L3 router~\cite{basic-routing}.
Using symbolic execution, we find input packets for which the program
fails to specify the egress specification, leading to unspecified behavior.

To do so, we prepopulate the tables with entries from the unit test provided
along with the program.
We then simply start the program with a single symbolic packet ($P$)
from a symbolic port in the input packet stream (the \cellk{in} cell).
Our goal is to find an input packet that leads the program to a state in
which neither packet is dropped, nor its egress specification is set.
We run the program in the (symbolic) search mode.
The search returns multiple inputs which can lead to undefined egress specification.
Here we only discuss one of the more interesting ones:
the search result suggests that if
"P has ethernet as its first header and ethernet.etherType != 0x0800",
then the program will end up with an undefined egress.

\begin{figure}[h]
\vspace*{-2ex}
  \centering
  \begin{lstlisting}[language=p4,basicstyle=\footnotesize\ttfamily]
  ...
  parser start {return parse_ethernet;}
  parser parse_ethernet {
    extract(ethernet);
    return select(latest.etherType) {
      0x0800 : parse_ipv4;
      default: ingress;
    }}
  control ingress { if (valid(ipv4)) { ... } }
  \end{lstlisting}
  \vspace*{-2ex}
    \caption{Part of a basic L3 router~\cite{basic-routing}}
    \label{fig:basic-routing}
  \vspace*{-3ex}
\end{figure}
Figure~\ref{fig:basic-routing} shows the relevant snippet of the program.
A simple manual inspection confirms the finding.
The parser extracts the \pcode{ethernet} header and
checks \pcode{etherType}.
If it is equal to 0x0800 (i.e the IPv4 ether type),
the parser then proceeds to extracting the \pcode{ipv4} header (not shown).
Otherwise, instead of, say, dropping the packet, the program starts
the ingress pipeline.
At the beginning of the ingress, the program checks the validity of
the \pcode{ipv4} header.
If valid, the pipeline applies a sequence of tables that may
set the egress specification (not shown).
Otherwise the program does not apply any tables and the
egress specification remains undefined.
Thus, under the given constraints, packet is not dropped,
\pcode{ipv4} is invalid, and the egress specification is undefined.

\subsubsection{Data Plane Verification}
\label{sec:dataplane-verification}
There is a growing interest towards \textit{data plane verification} tools
such as~\cite{hsa,netplumber,veriflow,deltanet,apv}.
These tools analyze the table entries in a snapshot of the data plane
and look for violation of properties of interest.
The verification of these properties usually requires answer
to queries of the following form: \textit{What kind of packets from node A will reach node B}?
While using various smart ideas to achieve better performance,
all these tool are based on the same basic idea:
symbolic reasoning over the space of packet headers.

Using our semantics, we can answer such queries by inserting
a symbolic packet at, say, node $A$ and using symbolic execution to find
the constraints on the packets that end up at node $B$.
The tools mentioned above use simplified hardcoded/adhoc models of
packet processors in their analysis
and miss the internal details of such devices.
They need to be re-engineered to change their model of packet processors.
There is no such need in our case.
Moreover, as will be shown in the next section, these tools can verify a very restricted
class of properties.
We eliminate these limitations.



\subsection{Program Verification}
\label{sec:deductive-verification}


K features a language independent program verification infrastructure
based on Reachability Logic~\cite{kprover}.
It can be instantiated with the semantics of a programming language such as P4
to automatically provide a sound and relatively complete program verifier
for that language.
In this system, properties to be verified  are given
using a set of reachability assertions, where
each reachability assertion is written as a rewrite rule.
%
A reachability assertion asserts that starting from any configuration
matching the left hand side of the assertion,
by execution using the input semantics, one will either eventually
reach a configuration that matches the right hand side of the assertion
or never terminate.

The standard pre/post conditions and loop invariants used in Hoare style
program verification can be encoded as reachability assertions.
Intuitively, a Hoare triple $\{P\}C\{Q\}$ becomes
"$C \land P$ \textit{rewrites to} $. \land Q$"  where ``.'' is
the empty program~\cite{kevm}.

\subsubsection{The Load Balancer Program}
To showcase the use of the program verifier,
we provide a simple P4 program and verify a simple property about it.
\begin{figure}[h]
\vspace*{-2ex}
  \centering
  \begin{lstlisting}[language=p4,basicstyle=\footnotesize\ttfamily]
  header_type meta_t { fields { reg_val : 8; } }
  metadata meta_t meta;
  parser start{ return ingress; }
  register reg { width: 8; instance_count: 1; }
  action read_reg(){
    register_read(meta.reg_val,reg,0); }
  table read_reg_table{
    reads{ meta.valid : exact; }
    actions{ read_reg; } }
  action balance(port,val){
    modify_field(standard_metadata.egress_spec,port);
    register_write(reg, 0, val);}
  table balance_table{
    reads{ meta.reg_val : exact; }
    actions{ balance; } }
  control ingress{ apply(read_reg_table); apply(balance_table); }
  \end{lstlisting}
\vspace*{-2ex}
  \caption{A simple load balancer}
  \label{fig:load-balancer}
\vspace*{-2ex}
\end{figure}
The program above is meant to balance its incoming packets
(from any port) between two output ports.
This is done using a register whose value alternates between 0 and 1
across incoming packets.
The program features a single register, a metadata instance, and two tables.
The parser starts ingress without extracting anything.
We install a single entry in \pcode{read\_reg\_table} to call
action \pcode{read\_reg}.
The action copies the register value (at index 0) into \pcode{meta.reg\_val}
\footnote{This is done because register values can not directly be matched in tables.}.
We install two rules in \texttt{balance\_table}.
One rule matches if \pcode{meta.reg\_val = 1} and calls \pcode{balance(1,0)}.
The other rule matches if \pcode{meta.reg\_val = 0} and calls \pcode{balance(0,1)}.
\pcode{balance(p,v)} modifies the register (at index 0) with value \pcode{v}
and effectively sends the packet to port \pcode{p}.
Our goal is to prove that this program (along with its table entries) correctly balances the load.
Specifically, we want to prove the following:

\textit{Property}: For any input stream of packets,
after processing all the packets,
no packet is dropped and no new packet is added;
all the packets in the output are either sent to port 0 or port 1;
and the absolute difference between the number of packets sent
to ports 0 and 1 is less than or equal to 1.
Albeit simple, none of the data plane verification tools mentioned
above are capable of proving it.
They lack either support for stateful data plane elements or
support for reasoning over an unbounded (i.e symbolic) stream
of packets.

In K, the property is captured by the following reachability assertion.
For presentation purposes we have omitted the less relevant,
mostly static parts of the specification which hold
the program and the table entries.
The full specification can be found in~\cite{p4k}.
\vspace*{-2ex}

{\footnotesize
\kequation{spec}{
\begin{array}{@{}l@{}}
\kall{k}{\reduce{@execute}{@end}} \mathrel{}
\kall{stateful}{
      \kall{name}{reg} \mathrel{}
      \kall{vals}{0\mapsto{\reduce{0}{\_}}} \mathrel{}
}
\kall{in}{\reduce{\variable{I}}{.}}
\\ \mathrel{}
\kall{out}{\reduce{.}{\variable{?O}}}
\hspace{1px} \text{\ \ ensures} \hspace{1px}
    \ensuremath{
    \begin{array}{l}
        \|onPort(?O,0) - onPort(?O,1))\| \leq 1 \\
        \land onPort(?O,0) + onPort(?O,1)) =  size(?O) \\
        \land \hspace{1px} size(?O) = size(I)
    \end{array}
    }
\end{array}
}
}
In the specification above, \pcode{@excute} is the program state right
before the execution starts.
\pcode{@end} is state after all the input packets are
processed\footnote{This state is only added due to technical reasons for
verification purposes, as actual P4 programs never terminate.
We add a rule causing the program to jump to this state once the input packet
stream becomes empty.}.
$I$ is a (universal) symbolic variable representing the input packet stream
and $?O$ is an existential symbolic variable representing the output stream.
Symbol ``$.$'' in both \cellk{in} and \cellk{out} cells represents
an empty packet stream.
The rewrite in the \cellk{vals} cell says that the value of the register
\texttt{reg} (at index 0) is 0 at start\footnote{We made the
assumption that registers are initialized to 0.},
and its value at the end is not relevant to the assertion.
The keyword \pcode{ensures} adds logical constraints on the right hand
side of the assertion (i.e the post condition).
Function \pcode{onPort(s,p)} returns the the number of packets in stream $s$
belonging to port $p$.
Function \pcode{size(s)} returns the length of stream $s$.

Our semantics of P4 contains a main loop over the stream of input packets.
Since in our property the input is a symbolic list with an unbounded length,
similar to Hoare logic, we need a loop invariant.
To prove our property, we provide the loop invariant as the
following reachability assertion:

{\footnotesize
\kequation{spec}{
\begin{array}{@{}l@{}}
\\[-6ex]
\kall{k}{\reduce{@nextPacket}{@end}} \mathrel{}
\kall{stateful}{
      \kall{name}{reg} \mathrel{}
      \kall{vals}{0\mapsto{\reduce{\variable{R}}{\_}}} \mathrel{}
}
\kall{in}{\reduce{\variable{I}}{.}}
\\ \mathrel{}
\kall{out}{\reduce{O1}{\variable{?O2}}}
\hspace{1px} \text{requires} \hspace{1px}
    \ensuremath{
    \begin{array}{l}
     (R = 1 \land onPort(O1,0) = onPort(O1,1)) + 1\lor \\
     R = 0 \land onPort(O1,0) = onPort(O1,1))) \\
     \land onPort(O1, 0) + onPort(O1, 1) = size(O1)
    \end{array}
    }
\\ \mathrel{}
\hspace{1px} \text{ensures} \hspace{1px}
    \ensuremath{
    \begin{array}{l}
     \|onPort(?O2,0) - onPort(?O2,1))\| \leq 1 \\
     \land onPort(?O2, 0) + onPort(?O2, 1) = size(?O2) \\
     \land size(I) + size(O1) = size(?O2)
    \end{array}
    }
\end{array}
}
}
Here, \pcode{@nextPacket} is the head of the main loop over the input packet stream.
Keyword \pcode{requires} puts logical constraints on the left hand side
of the assertion (i.e the precondition).
The assertion reads as:
starting from the head of the main loop,
given the constraints in \pcode{requires} are satisfied,
if the program terminates,
it will reach an \pcode{@end} state that satisfies the constraints in \pcode{ensures}.


We gave the two assertions to the K's program verifier instantiated with
our P4 semantics.
The verifier successfully proved the loop invariant and the first reachability
assertion (i.e., the desired property).
The verification took about 80s.

In this example, we used concrete table entries as the entries are part of the
functionality that we aimed to verify.
In general, depending on the property, the tables -- as well as anything else -- can
be symbolic.
We show an example in the next section.







\subsection{Translation Validation}
\label{sec:translation-validation}
P4 programs eventually need to be compiled into the instruction set (i.e the
language) of the target hardware for execution.
With any compilation, there is the question of whether or not the semantics of
the input program is preserved by the compiler.
Currently the compilers usually lack formal semantic preservation guarantees
since providing such guarantees requires a significant effort.
The issue is even more pronounced when sophisticated compiler optimizations are
involved.
A promising alternative approach is to verify each instance of compilation
instead of the whole compiler.
This approach, known as \textit{tranlation validation}~\cite{tv}, aims to verify
the semantic equivalence of a program and its compiled counterpart, potentially
using hints from the compiler.

Recently, K has introduced a prototype tool (named KEQ~\cite{k-eq}) for
cross language program equivalence checking using a generalized
notion of bisimulation.
The notion enables us to mask irrelevant intermediate states and
consider only the relevant states in comparing two program executions.
KEQ takes the K semantics of the two programming languages,
two input programs written in the respective languages,
and a set of synchronization points as input, and
checks whether or not the two programs are equivalent.

Each synchronization point is a pair of symbolic states (called \textit{cut}s)
in the two input programs.
The meaning of synchronization is defined by the user as a logical constraint over
the given pair of symbolic states.
It usually consists of checking the equality of certain relevant values.
Each cut in the pair is essentially a pattern over the configurations
of the semantics of the respective languages
(similar to the right or left hand side of rewrite rules).
The user labels one or more synchronization points as \textit{trusted}.
These points are assumed to already be bisimilar.
Usually one (and the only one) such point is the end of the two programs
and the constraint is the equality of the respective output values.

For the rest of the synchronization points, the equivalence checker checks
whether the given points are bisimilar.
It basically means that starting from the two cuts in a synchronization point,
using the semantics of the respective languages, all reachable
synchronization points are respectively bisimilar.
Normally one such point is the start of the two programs.
The constraint is the equality of the respective input values.
Additional synchronization points may be needed as well, such as the
beginning of unbounded loops.
We refer the interested readers to~\cite{k-eq} for more details on KEQ.

\subsubsection{P4 $\to$ IMP+ Translation Validation}
We illustrate KEQ through a small example.
We check the equivalence of a simple P4 program with a program written in another language.
For this purpose, we developed a very simple imperative language called IMP+.
The language syntactically resembles C,
although semantically it is much simpler.
We also developed the semantics of IMP+ in K.
We provide a set of API functions for the language to send and receive packets, read tables, etc.
The name of these functions are prefixed with the ``\#'' symbol.
For simplicity, we directly provide semantics to such functions in our semantics.
We chose the simple P4 program in Figure~\ref{fig:simple-p4-program} for translation.
We manually translate it into IMP+ as
follows:\footnote{We assume packets with undefined egress specification will be dropped.}
\begin{lstlisting}[basicstyle=\footnotesize\ttfamily]
int h1_f1; int h1_f2; bool h1_valid;
int sm_egress_spec;
bool parse(){ return start(); }
bool start(){
  if (! #has_next(8)){ return false; }
  h1_f1 = #extract_next(8, false);
  if (! #has_next(8)){ return false;}
  h1_f2 = #extract_next(8, false);
  h1_valid = true;
  return true;}
void a(int n){
  h1_f2 = n; sm_egress_spec = 1; }
void b(){ sm_egress_spec = 2; }
void apply_t(){
  //p2
  while (#get_next_entry()) {
    if (#entry_matches(h1_f1)){
      #call_entry_action(); return; }}
  if (#has_default_action()){
    #call_default_action(); }}
bool process_packet(){
  #reset();
  sm_egress_spec = -1;
  h1_valid = false;
  if (! parse()){ return false; }
  if (sm_egress_spec == -1){return false;}
  return true;}
void deparse(){
  #emit(h1_f1); #emit(h1_f2); #add_payload(); }
void main(){ //p0
  //p1
  while (#get_next_packet()){
    if (!process_packet()){ #drop(); }
    else{ deparse(); #output_packet();}}}
//p3 [trusted]
\end{lstlisting}
The goal is to prove the equivalence of the two programs.
Our notion of equivalence is defined as follows:
for any input stream of packets,
and for any table entries in table $t$,
at the end of processing all the input packets,
the two programs generate the same output stream of packets.
To do so, we manually provide a few synchronization points.
We have annotated the IMP+ program with the points.
Next we informally describe the points and their constraints.
The full specification can be found in~\cite{p4k}.

\begin{figure}[t]
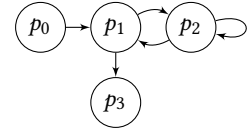

  \tikz {
    \tikzset{vertex/.style = {shape=circle,draw}}
    \tikzset{edge/.style = {->,> = latex'}}

    \node[vertex] (p0) at  (-1,1) {$p_0$};
    \node[vertex] (p1) at  (0,1) {$p_1$};
    \node[vertex] (p2) at  (1,1) {$p_2$};
    \node[vertex] (p3) at  (0,0) {$p_3$};
    \draw[edge] (p0) to (p1);
    \draw[edge] (p1) to[bend left] (p2);
    \draw[edge] (p2) to[bend left] (p1);
    \draw[edge] (p2) to[loop right] (p2);
    \draw[edge] (p1) to (p3);
  }
  \vspace{-2ex}
  \caption{Abstract transition relation between $p_0 ... p_3$.}
  \label{fig:sync-points}
  \vspace{-4ex}
\end{figure}

$p_0$ is the start of the two programs.
The condition associated with this points is the equality
of the respective input streams, table entries, and table default actions.

$p_1$ is the main loop over the input packets.
Its condition is same as $p_0$'s condition plus the equality of the respective
current output packet streams.

$p_2$ is the loop over table entries.
The condition is $p_1$'s condition plus
the equality of the field values in the parsed representation of the P4 program
and the corresponding variables in the IMP+ program,
plus the equality of the index of iteration of the tables
entries\footnote{Note that in our semantics of both P4 and IMP+, the table entries
are sorted in the descending order of their priority.}, and the equality of the current packet payloads.

$p_3$ is the end of execution\footnote{For technical reasons we assume both programs
terminate once the input packet stream becomes empty.}.
The condition is the equality of the respective output packet streams.
$p_3$ is \textit{trusted}.



Figure~\ref{fig:sync-points} illustrates the abstract transition relation between
the points.
Each arrow represents multiple rewrite steps in each program, ignoring the
irrelevant (possibly non-equivalent) intermediate states of the programs.
Note how this abstraction enables us to establish the equivalence even though
the programs are written in two quite different programming languages.

Given the synchronization points, KEQ was able to prove the equivalence.
Although the program is very simple, we believe it captures the essence of
many of the programs that are used in practice.
In addition, note that we provided the synchronization points by hand.
In practice, the compiler can automatically provide this information
as it has enough knowledge during the translation.




%% file: sections/related_work.tex
\section{Related Work}
\label{sec:related_work}
\subsection{Semantics of Programming Languages in K}
The K framework has been used to provide complete executable semantics for
several programming languages.
Here we briefly overview the more relevant work.

KCC~\cite{kcc} formalizes the semantics of C11, passing
99.2\% of GCC torture test suit, more than what the
GCC and Clang could pass.
Later work~\cite{kcc-undefined} develops the "negative" semantics of C11
and is able to identify programs with undefined behavior.
In our work, we identify unspecified behavior by lack of semantics.

K-Java~\cite{kjava} formalizes Java 1.4 and follows a
\textit{test-driven methodology} to manually provide a suite of 800+ tests.

KJS~\cite{kjs} provides semantics for JavaScript passing all 2700+ tests
in a conformance test suite~\cite{ecma}.
The authors introduce the notion of
semantic coverage for test suites which has inspired our work.

In these works, the language design predates the formalization effort
by several years.
Consequently, although more complex, these languages are quite stable.
P4 is still at the early stages of the language design process
and is relatively unstable.
This made the formalization effort challenging.


Recently, KEVM~\cite{kevm} formalizes the Etherium Virtual Machine~\cite{evm},
successfully passing a suite of 40K official tests.
Like P4K, KEVM targets a new language and reveals problems in its specification.

These works (except K-Java) rely heavily on existing tests to provide semantics.
In our case, such a comprehensive test suite still does not exist.
The only test suite that we found covers less than 54\% of our semantics.


\subsection{Network Verification}
Bugs happen frequently in networks and lead to performance problems,
service disruptions, security vulnerabilities, etc.
Scale and complexity of networks make answering even the simplest
functional correctness queries prohibitively hard to answer manually.
This has ignited research into automating the process
of \textit{network verification} which can be broadly categorized as follows:

\textit{Data plane verification} reactively checks network wide
properties in a data plane by analyzing snapshots of it.
We have already discussed the work in this area in Section~\ref{sec:applications}.
None of these tools readily supports P4.
An exception is~\cite{p4nod} which will be discussed in the next section.



\textit{Control plane verification} proactively ensures a network is free
of latent bugs by analyzing its control plane logic.
The literature targets both traditional and SDN control planes.
The techniques include
static analysis (e.g.~\cite{rcc}),
simulation and emulation (e.g.~\cite{batfish, crystalnet}),
model checking (e.g.~\cite{plankton}),
SMT solving (e.g.~\cite{minesweeper}),
testing (e.g.~\cite{nice, flowtest}),
control plane abstractions (e.g.~\cite{era, arc}),
and deductive verification (e.g.~\cite{vericon}).
All of these works (except~\cite{crystalnet}) assume a fixed and simple model
for the data plane elements which misses the internal details of the devices.

Although control plane is out of the scope of this work, it it worth mentioning
that SDN controllers are usually written using mainstream programming languages
\footnote{Though there is a recent progress towards high level
programming languages for controllers~\cite{netkat, temporalnetkat, snap}.}
for which K semantics exist (e.g. ONOS~\cite{onos} is in Java and
P4Runtime~\cite{p4runtime} is in C).
By combining the K semantics of controller programs with our semantics of
P4 data planes, we can analyze a complete model of the whole network.
We leave this as future work.






\subsection{Semantics and Analysis of P4}
\label{sec:p4-analysis}
We are not aware of any extensive efforts to formalize the P4 language.
P4NOD~\cite{p4nod} (on paper) provides a big step operational semantics of
a subset of the P4 language.
The authors use the semantics to provide a translator from P4 to Datalog.
The result is used in P4 data plane verification using a Datalog engine
optimized for this purpose~\cite{nod}.
The authors also use the tool to catch a class of bugs, called well-formedness bugs,
that are unique to P4 networks.
Finally, the authors show an example of P4 to P4 equivalence check.

The primary focus of our work is the language itself and its problems.
We provide a modular small step operational semantics for all features of P4.
Using K tools, among other things, we too are able to perform data
plane verification and detect well-formedness bugs.
We have also shown an example of translation validation between P4 and other
languages defined in K.
In P4NOD the trust base consists of the semantics,
the Datalog engine, and the P4 to Datalog translator.
In P4K, it consists of the semantics and the K framework.
We leave a quantitative comparison of the two works as a future work.





%% file: sections/conclusion.tex
\section{Conclusion, Discussion, Future Work}
\label{sec:discussion}
\label{sec:conclusion}
We have presented P4K, the first complete semantics of \pft.
Through our formalization process, we have identified many problems
with the language specification.
We automatically provide a suite of analysis tools derived directly from our semantics.
We have discussed and demonstrated the applications of some of the tools for
P4 developers and designers.


With the introduction of \pst, \pft may sooner or later be deprecated,
especially because \pst addresses many of \pft's issues through
backwards-incompatible changes.
Nevertheless, we think that formalizing \pft was a worthwhile effort.
There are still important applications written
in \pft (e.g.~\cite{switch-p4}) that do not have a \pst equivalent.
The language consortium provides a translator from \pft to \pst.
However, without a clear semantics of \pft, the translation itself
might be problematic.
We are aware of at least one instance~\cite{p4-14-16-bmv2-header-stacks}
in which the translator's \pst output is not equivalent to its \pft input.

We plan to formalize \pst in near future.
We believe transition to \pst will be straight forward.
\pst has actually a smaller core language compared to \pft.

Beside other future directions discussed throughout the paper,
we also plan to use our semantics to analyze
real world P4 programs (specially~\cite{switch-p4, p4paxos}) and networks.

Another interesting use case of our semantics which we leave as a future work
is to automatically generate a test suite that covers all of our semantic rules.
Such a test suite can be very useful for P4 compiler developers, because they can
regenerate the tests each time the language changes.

We conclude by a lesson we learned.
It is relatively easy and extremely beneficial to rapidly apply formal methods
at the early stages of a language design process.
Not only it helps the designers quickly identify problems in
their design and produce more robust languages,
but also (in case a framework like K is employed) it can save
time and effort by automatically providing various useful tools
for the language.





%% file: sections/acknowledgments.tex
\section*{Acknowledgments}
The authors would like to thank the K development team (specially Daejun Park)
for their help with the K framework.
We would also like to thank Nate Foster and the members of P4 Language
Consortium for their support.
We thank Brighten Godfrey, Farnaz Jahanbakhsh, and Alex Horn for their feedback.
This material is based upon work supported by the National Science Foundation under
Grant No. CCF-1421575.